\documentclass[reprint,superscriptaddress,
 amsmath,amssymb,
pra,
]{revtex4-1}
\usepackage[utf8]{inputenc}

\usepackage{graphicx}
\usepackage{dcolumn}
\usepackage{bm}
\usepackage{hyperref}
\usepackage{ulem}
\usepackage{xcolor}
\usepackage{amsmath, amssymb}
\usepackage[T1]{fontenc}

\begin{document}

\preprint{APS/123-QED}

\title{Finite-range effects in the two-dimensional repulsive Fermi polaron}

\author{Raúl Bombín}
\email{raul.bombin@ehu.eus}
\affiliation{Departamento de Física de Materiales, Facultad de Químicas, UPV/EHU, Apartado 1072, 20080 San Sebastián, Spain}
\affiliation{Centro de Fısica de Materiales Centro Mixto CSIC-UPV/EHU, Edificio Korta, Avenida de Tolosa 72, 20018 San Sebastián, Spain}
\affiliation{Departament de F\'{i}sica, Universitat Polit\`{e}cnica de Catalunya, Campus Nord B4-B5, E-08034, Barcelona, Spain}

\author{Viktor Cikojevi{\'c}}
\email{viktor.cikojevic@pmfst.hr}
\affiliation{Departament de F\'{i}sica, Universitat Polit\`{e}cnica de Catalunya, Campus Nord B4-B5, E-08034, Barcelona, Spain}
\affiliation{University of Split, Faculty of Science, Ru\dj era Bo\v{s}kovi\'ca 33, HR-21000 Split, Croatia}

\author{Juan Sánchez-Baena}
\email{juan.sanchez.baena@upc.edu}
\affiliation{Departament de F\'{i}sica, Universitat Polit\`{e}cnica de Catalunya, Campus Nord B4-B5, E-08034, Barcelona, Spain}

\author{Jordi Boronat}
\email{jordi.boronat@upc.edu}
\affiliation{Departament de F\'{i}sica, Universitat Polit\`{e}cnica de Catalunya, Campus Nord B4-B5, E-08034, Barcelona, Spain}

\date{\today}

\begin{abstract}
We study the repulsive Fermi polaron in a two-component, two-dimensional 
system of fermionic atoms inspired by the results of a recent experiment with $^{173}$Yb atoms [N. Darkwah Oppong \textit{et al.}, Phys. Rev. Lett. \textbf{122}, 193604 (2019)]. We use the diffusion Monte Carlo method to report properties such as the polaron energy and the quasi-particle residue that have been measured in that experiment. To provide insight into the quasi-particle character of the problem, we also report results for the effective mass. We show that the effective range, together with the scattering length, is needed in order to reproduce the experimental results. Using different model potentials for the interaction between the Fermi sea and the impurity, we show that it is possible to establish a regime of universality, in terms of these two parameters, that includes the whole experimental regime. This illustrates the relevance of quantum fluctuations and beyond mean-field effects to correctly describe the Fermi polaron problem.
\end{abstract}


\maketitle


The problem of a single impurity surrounded by a medium has been studied in several quantum many-body systems since its initial formulation \cite{Landau1933,Landau1948}. For decades the impurity problem has been studied in many fields: from condensed matter (\textit{cf}. ~\cite{Edwards1992,BoronatHNC, Boronat1999}, for example, for studies in helium), to neutron matter~\cite{Bishop1970}, and, more recently in atomically thin semiconductors \cite{Sidler2017}.
In some particular systems, the presence of impurities is crucial to explain some of its physical properties, e.g., the  Kondo effect, originated by the presence of magnetic impurities~\cite{kondo64}, and of Anderson's orthogonality catastrophe~\cite{anderson67}, in fermionic systems.  Under some physical conditions, the impurity coupled to the medium behaves as a quasi-particle (polaron), whose properties differ drastically to those of the impurity. For example, in the case where the medium is fermionic, it is called Fermi polaron~\cite{massignan14,Ngampruetikorn2012}

Ultracold atoms, due to their high tunnability, constitute an excellent platform for the investigation of the Fermi polaron. Precisely for this, they have boost intense experimental and theoretical work in different geometries that include one \cite{Wenz2013,Gharashi2015,Mistakidis2019,Song2019}, two (2D) \cite{Bombin2019,Darkwah2019,adlong2020quasiparticle} three dimensional (3D) \cite{scazza17,adlong2020quasiparticle} configurations. Incidentally, efforts have not been restricted to the study of the ground state and  recently the study has been extended to finite temperature. In particular a crossover from the quantum Fermi polaron regime at low temperature to the classical Boltzmann as temperature increases has been reported by means of radio-frequency spectroscopy \cite{Yan2019_2} (\textit{cf}. for theoretical works \cite{Tajima2018,Tajima2019}).

Interestingly, the interaction between the impurity and the bath can be controlled by means of a Feshbach Resonance. 
This has allowed to experimentally access
the polaron physics in different regimes employing two-
component mixtures of ultracold gases with a very small
concentration of one of the components . This includes mixtures of two different hyper-fine levels of the same atomic specie~\cite{Jorgensen2016} and of different atoms ~\cite{Tempere2009,Baarsma2012,Cucchietti2006,Hu2016,Yan2019}. While initially only alkali atoms were employed~\cite{nascimbene09,schirotzek09,kohstall12,koschorreck12,ong15,scazza17}, the recent discovery of Orbital Feshbach resonance (OFR) in $^{173}$Yb~\cite{Zhang2015,Hofer2015,Pagano2015} has made it possible to study new physical phenomena in the quantum degenerate regime, for example as spin exchange and SU(N)-symmetric collisions \cite{Gorshkov2010,Scazza2014,Zhang1467}. This has motivated to spectroscopically probe the energies of the repulsive and attractive branches of the polaron, and to measure the quasi-particle residue by driving Rabi oscillations (both in 2D ~\cite{Darkwah2019} and 3D systems \cite{scazza17,Deng2018}). 

Previous diffusion Monte Carlo (DMC) calculations reveal the existence of a universal regime in terms of the  gas parameter $na_s^2$ for the 2D repulsive Fermi polaron, that stands for values $n a_s^2 \le 10^{-3}$~\cite{Bombin2019}, with $n$ the particle density and $a_s$ the 2D s-wave scattering length. These results contrast with the ones reported for unpolarized 2D, two-component Fermi system where the universal regime stands for $n a_s^2 \lesssim 10^{-2}$~\cite{comparin19}. This reflects the enhanced relevance of quantum fluctuations in 2D Fermi polaron problem \cite{Vliettinck2014}, being a good testbed to study the effect of impurity-bath correlations \cite{Kwasniok2020}.

Remarkably, the energy of the 2D polaron and the quasi-particle residue have been measured recently outside of the universal regime by exploiting the OFR in $^{173}$Yb~\cite{Darkwah2019}. As a consequence, mean-field theory and its first perturbative correction (the Lee-Huang-Yang (LHY) term),  which are functions of only the scattering length, are not sufficient to accurately describe this problem. Thus, to provide a quantitative description of the experimental results of Ref.~\cite{Darkwah2019}, effective range effects, which go beyond the LHY correction, must be taken into account. Here we study the repulsive 2D Fermi polaron in terms of the Fermi momentum times the scattering length $k_\mathrm{F}a_s$, that has a direct relationship with the gas parameter $k_\mathrm{F}a_s= \sqrt{4\pi n a_s^2}$, well outside the unitary limit. (For a DMC study of the 2D Bose polaron \textit{cf}. Ref. \cite{Ardila2020}). We restrict ourselves to a two-dimensional system to mimic the conditions of the experiment in Ref.~\cite{Darkwah2019}.

With the aim of extending the regime of universality, in this Letter we study the 2D Fermi polaron problem with a model in which the two-body potential is constructed taking into account both  the s-wave scattering length and the effective range.
We fix both quantities to values compatible with the experiment of Ref~\cite{Darkwah2019}. We perform our calculations with two different model potentials, detailed below, in order to check if the scattering length and the effective range are sufficient to quantitatively describe the system in the regime of $k_\mathrm{F}a_s$ considered. A similar approach has been carried out in other systems, where good agreement with experimental results has been found, see for example Refs.~\cite{Cikojevic2019,Cikojevic2020,cikojevi2020qmcbased} for results regarding quantum droplets in Bose-Bose mixtures. Our results show that the explicit consideration of the finite range of the interaction allows for an excellent agreement with experimental data. Interestingly, we observe a regime of universality in terms of the two scattering parameters that covers the experimental regime of Ref. \cite{Darkwah2019} for the 2D Fermi polaron. 

Our microscopic approach is based on the DMC method.
DMC allows to accurately describe the ground state of quantum systems both in the dilute and in the strongly correlated regimes. Starting from a variational ansatz,
the initial wave function $\Psi_T(\textbf{R})$ is propagated in imaginary time
keeping its nodal surface constant. In this way, one keeps the Fermi sign problem under control, leading to a statistical representation of the best possible wave function within a nodal surface
constraint (fixed-node approximation (FN)~\cite{Reynolds1982}). As a consequence, FN-DMC produces variational results whose quality is related to the accuracy of the model nodal surface. The trial wave function, is taken as the usual Jastrow-Slater ansatz, consisting of a product of a Jastrow factor $\Psi_J(\textbf{R})$, that is symmetric to the exchange of particles, times an anti-symmetric part $\Psi_A(\textbf{R})$ that is taken to be a Slater determinant of plane waves, which is accurate enough for the low  densities considered in this work~\cite{Comparin2018}. Here, $\mathbf{R} = \lbrace \mathbf{r}_1, \dots,
\mathbf{r}_{N_\uparrow}, \mathbf{r}_\downarrow \rbrace$ is the set of all
$N_\uparrow + 1$ particle coordinates, and
$\mathbf{R}_\uparrow$ accounts only for the coordinates of the $N_\uparrow$ particles of the bath.  The Jastrow term is chosen as a product of two-body correlation functions, which are taken as the ground-state wave function of the two-body problem at short distances matched with a phonon-like long-range term \cite{Bombin2019}. 
The propagator that we employ is accurate up to second order in the imaginary time step. Off-diagonal estimators have an additional bias related to the choice of $\Psi_T(\textbf{R})$. In order to correct this, up to first order in the trial wave function, we use the extrapolated estimator technique \cite{Bombin2019}. The error bars of our results  account for this source of error, statistical Monte Carlo uncertainty, the bias originated from the finite imaginary time step, and finite-system size effects.A more detailed description of the method and the trial wave function can be found in the Supplementary Material.

It is worth to remark that a similar attempt has been made recently with the aim of extending the current state of the art theory. In Ref.~\cite{adlong2020quasiparticle} the authors present a model that, by effectively including finite-range effects,  is able to reproduce the Rabi oscillations both from the 3D \cite{scazza17} and 2D \cite{Darkwah2019} Fermi polaron experiments. On the other hand, although the quasi-particle residue that they report improves previous theoretical results, it is not able to match the experimental ones. Besides that, some controlled numerical results (without considering effective range effects) have been presented in the literature for the Fermi polaron (\textit{cf}, for example, Ref.~ \cite{Goulko2016} for the resonant Fermi polaron and Refs.~ \cite{prokofev2008,Vlietinck2013,Kroiss2015,VanHoucke20} for diagrammatic Monte Carlo studies of the Fermi polaron.

To describe the 2D Fermi polaron, we study a system of  $N = N_{\uparrow} + 1$ particles composed by a Fermi sea of $N_{\uparrow}$ non-interacting fermions and a spin $\downarrow$  impurity, all with the same mass $m$. The Hamiltonian of the $N$-particle system reads
\begin{equation}
\hat{H} =
-\frac{\hbar^2}{2m}  \nabla_\downarrow^2
-\frac{\hbar^2}{2m} \sum_{i=1}^{N_{\uparrow}} \nabla_i^2 +
\sum_{j=1}^{N_{\uparrow}}V^{\mathrm{int}}(r_{\downarrow j}) \ ,
\label{eq:Hamiltonian}
\end{equation}
where $r_{\downarrow j} \equiv |\mathbf{r}_\downarrow - \mathbf{r}_j|$ is
the distance between a bath particle at $\mathbf{r}_j$ and the impurity position
$\mathbf{r}_\downarrow$. The two-body potential  $V^{\mathrm{int}}(r)$ models the interaction of the polaron with the bath. We use two different models: The first one is a Square-Well-Soft-Core (SWSC) potential, which reads
\begin{equation}
V^{\mathrm{int}}(r)=
\begin{cases}
-U_0 , &  0 < r < R_0 ,\\
U_1 , &  R_0 < r < R_1, \\
0 , &  R_1 < r < \infty \ ,\\
\end{cases}
\label{eqn:pot}
\end{equation} 
with all the parameters ($R_0$, $R_1$, $U_0$, $U_1$) being positive. The second model is a Soft-Core (SC) potential, that can be considered a limiting case of the previous one by setting $R_0$ and $U_0$ to zero, and thus it is uniquely described by $U_1$ and $R_1$.
The values of the parameters of Eq.~\eqref{eqn:pot} are reported in Table~\ref{table:potential_params}. These values are chosen so that the potential reproduces the experimental results from Ref. \cite{hofer15}, which corresponds to setting $a_{\rm 3D} = 1878 a_0$ and $r_{\rm 3D}^{\rm eff} = 216 a_0$, with $a_0$ the Bohr radius \cite{hofer2015observation, porsev2014longrange}. In Ref.~\cite{hofer15}, it is reported a second value for the effective range, which is smaller ($r_{\rm 3D}^{\rm eff} = 126 a_0$), but we checked with DMC that experimental data are only reproduced by choosing the larger value, $r_{\rm 3D}^{\rm eff} = 216 a_0$. See the Supplementary Material for further details concerning the interaction models employed.

\begin{table}[tb]
\begin{tabular}{|l | c | c | c | c |}
\hline
 &  $R_0[a_{\rm 3D}]$ & $R_1[a_{\rm 3D}]$ & $U_0[\hbar^2 / (m a_{\rm 3D}^2)]$ & $U_1[\hbar^2 / (m a_{\rm 3D}^2)]$ \\ \hline
 SC   & 0   & 2.40593     & 0     & 0.425291    \\ \hline
 SWSC & 0.91627   & 2.29069           & 0.62099     & 0.576351 \\ 
\hline
\end{tabular}
\caption{Parameters of the interaction potentials (SWSC, SC) that reproduce the experimental values of  $a_{\rm 3D}$  and $r_{\rm 3D}^{\rm eff}$. }
\label{table:potential_params}
\end{table}

One of the most relevant quantities in the study of the polaron is the polaron energy, which is in fact the chemical potential of the impurity. Within the DMC framework, it can be evaluated by means of the energy difference
\begin{equation}
\varepsilon_p = \left[E (N_\uparrow, 1) - E(N_\uparrow,0)\right]_V \ ,
\label{eq:ener_pol}
\end{equation}
where $E (N_\uparrow, 0)$ is the energy of the $N$-particle pure system and $E (N_\uparrow, 1)$ is the one obtained when the impurity is added, keeping the volume constant. Within the mean-field approximation, the polaron energy is given by
\begin{equation}
\varepsilon_{\mathrm{MF}} = \frac{4 \pi\hbar^2 n }{m\ln(c_0na_s^2)} 
\ ,
\label{eq:ener_polMF}
\end{equation}
with $c_0$ a free parameter~\cite{Pitaevskii2016} that 
is related to the energy scale of the system. Following previous works~\cite{Bertaina2013,Comparin2018,Bombin2019}, we fix it so that the choice for the energy scale corresponds
to that of the free Fermi system ($E_\mathrm{F} = \frac{\hbar^2 k_\mathrm{F}^2}{2m} = 2 \hbar^2 \pi n / m$), which results into $c_0=e^{2\gamma} \pi/2 \simeq 
4.98$, where $\gamma\simeq 0.577$ is the Euler's gamma constant. 

\begin{center}
\begin{figure}[t]
\includegraphics[width=1.02\linewidth]{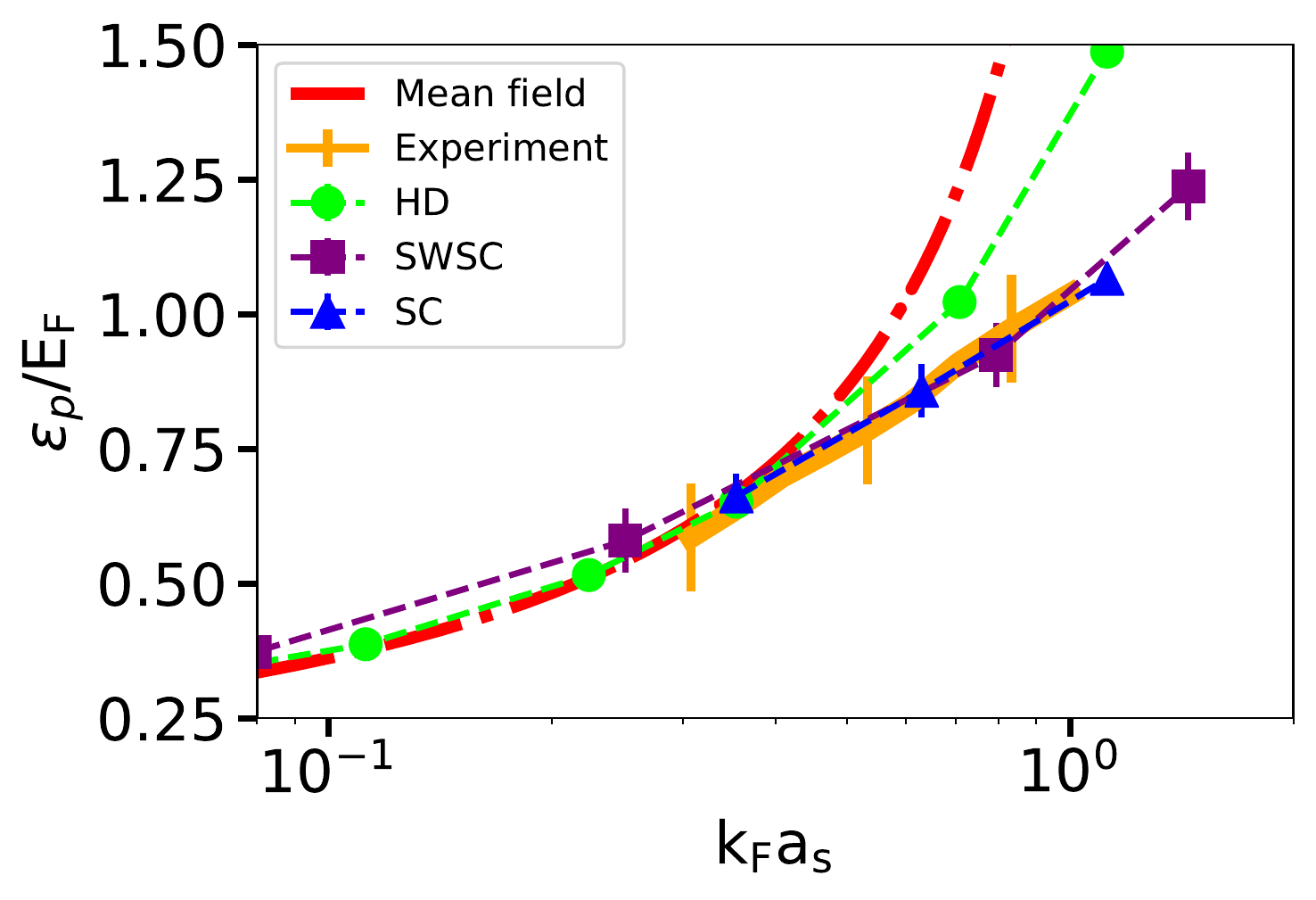}
\caption{Polaron energy in units of the bath Fermi energy $E_\mathrm{F} = 2 \hbar^2 \pi n / m$.
The red dashed line corresponds to the mean-field prediction of Eq.~\eqref{eq:ener_polMF} and the orange solid line  to the experimental results reported in Ref.~\cite{Darkwah2019}. Symbols are DMC results: circles correspond to the HD model reported in Ref.~\cite{Bombin2019}, and squares and triangles to results obtained using the SWSC and SC potentials, respectively. Dashed lines joining the symbols are guides to the eye.}
\label{fig:polaron_ener}
\end{figure}
\end{center}

In order to benchmark the present results, we compare our DMC energies for the SWSC and SC models to those obtained with a hard-disk  (HD) potential (from Ref.~\cite{Bombin2019}) and to the mean-field prediction (\ref{eq:ener_polMF}). While the hard-disk potential shares the same scattering length with the SWSC and SC potentials, its effective range is different and cannot be imposed in the construction of the model, as the HD potential has only one free parameter: the diameter of the disk.

We report in Fig.~\ref{fig:polaron_ener} our DMC results for the polaron energy corresponding to the SWSC and SC potentials. In the same figure, we include the mean field prediction, the HD model results \cite{Bombin2019}, and the experimental results of Ref.~\cite{Darkwah2019}.
As it can be seen in the plot, both mean-field theory and the HD model fail to reproduce the experimental data. As mentioned previously, mean-field theory can not accurately reproduce the experimental results because the system lays at gas parameters outside the universal regime. The HD model is also unable to provide the experimental energy because it only reproduces the experimental 3D scattering length, not the effective range. On the other hand, our two present models, in which both scattering length and effective range are fixed at the same time, show good agreement between them and with the experimental measurements~\cite{Darkwah2019}. Therefore, the dimensionless parameter $k_\mathrm{F}a_s$ is not the only relevant quantity to quantitatively describe the system and, to this end, finite range effects need to be included. Moreover, the independence on the specific shape of the potential, when both scattering length and effective range are reproduced, hints to a universal behavior in these two quantities for the range of $k_\mathrm{F}a_s$ values shown in the figure.

To better characterize the Fermi polaron, we evaluate properties that are related its quasi-particle character. First, we study the quasi-particle residue $Z$, which is defined as the overlap between the wave function of the system, featuring an interacting impurity, and the wave function of a pure system with a non-interacting impurity with zero momentum, $k=0$ \cite{Vlietinck2013}. Formally, it reads
\begin{equation}
Z = \left|\langle \Phi ^{\mathrm{NI}}|\phi\rangle\right|^2 \ .
\label{eqn:Z_def}
\end{equation}
For the calculations presented here, $\Phi ^{\mathrm{NI}}=|\mathrm{FS} + 1\rangle $, which
stands for a Fermi sea (FS) with an added non-interacting impurity with zero momentum. Following previous works~\cite{Punk2009,Guidini2015,Bombin2019}, we evaluate the quasi-particle residue from the long-range asymptotic behavior of the one-body density matrix when the interacting polaron moves in the Fermi bath.  In our DMC implementation, this is obtained from the following estimator
\begin{equation}
Z=\lim_{|\mathbf{r}^\prime_\downarrow-\mathbf{r}_\downarrow|\rightarrow L/2}
\left\langle \frac{\Psi_T(\mathbf{R}_{\uparrow},\mathbf{r}_{\downarrow}
^ { \prime } ) } { \Psi_T(\mathbf{R}_ { 
\uparrow},\mathbf{r}_{\downarrow})}
\right\rangle \ .
\label{eq.Z_residue}
\end{equation}

\begin{center}
\begin{figure}[t]
\includegraphics[width=1.02\linewidth]{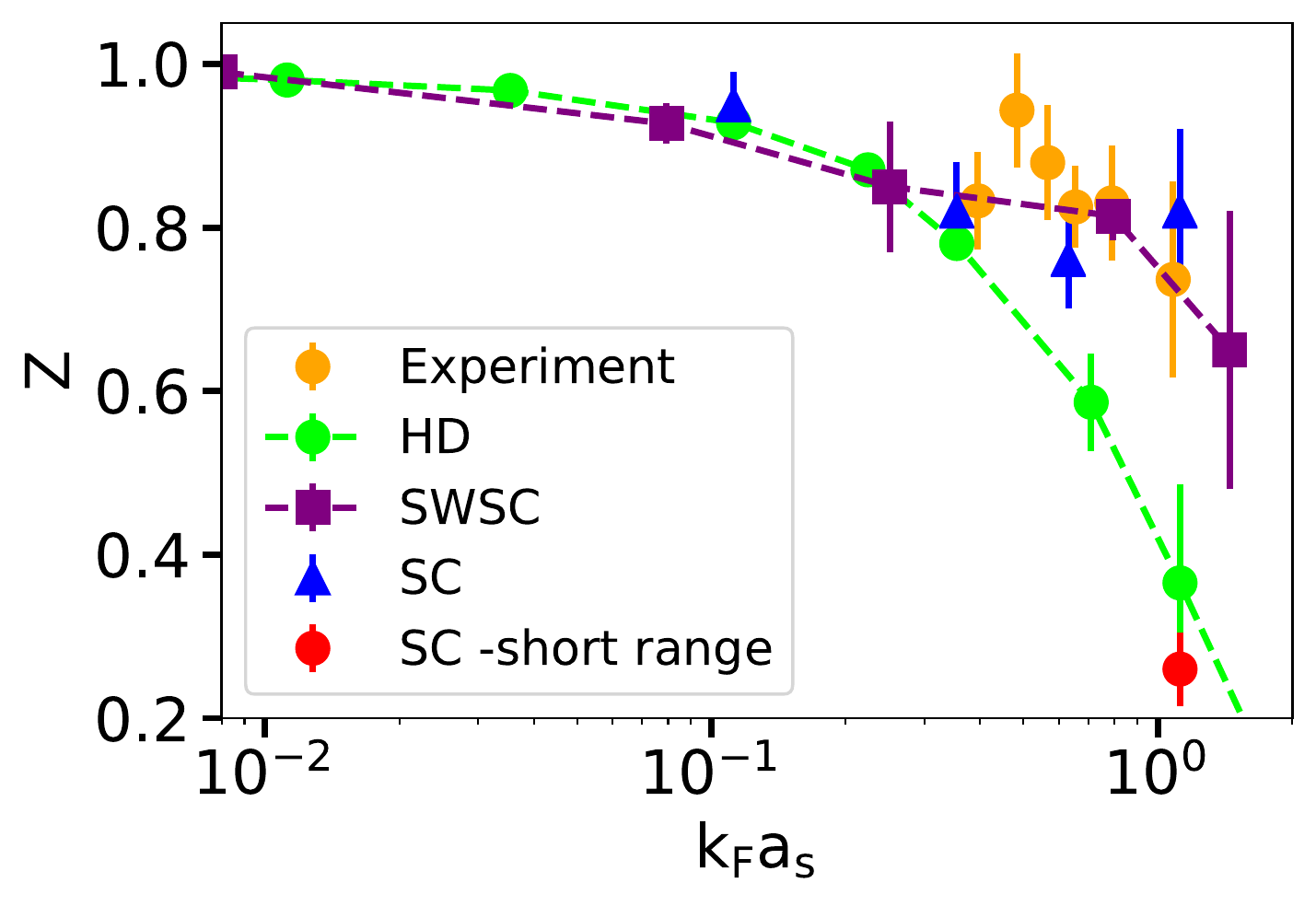}
\caption{Quasi-particle residue $Z$ as $k_\mathrm{F}a_s$.
Orange circles correspond to the experimental results reported in Ref. \cite{Darkwah2019}. Green circles, squares and triangles are results for the Hard-disk of Ref.~\cite{Bombin2019} and for the SWSC and SC models described in the text, respectively. The red point corresponds to a calculation with the soft disk potential with a small effective range. Dashed lines are guides to the eye.}
\label{fig:Z_residue}
\end{figure}
\end{center}

In Fig. \ref{fig:Z_residue}, we show DMC results for $Z$ using the SWSC, SC models. Again, we benchmark our results in the weakly interacting regime with a HD model. In the same figure, we include the experimental results of Ref~\cite{Darkwah2019} for this quantity. For values of $k_\mathrm{F}a_s < 0.2$, we find that the results of the three models coincide, in agreement to the universal regime for this observable, as reported in Ref.~\cite{Bombin2019}. However, as $k_\mathrm{F}a_s$ is increased, the results of the  SWSC and SC models, for which the effective range and scattering length are fixed simultaneously, show good agreement with the experimental data of Ref~\cite{Darkwah2019} and a clear discrepancy with the HD short-range model. These results reinforce the ones for the polaron energy and remark the importance of going beyond the usual mean-field prescription in order to provide an accurate description of the 2D Fermi polaron. For completeness, in the same plot, we also include a point in which the quasi-particle residue is evaluated for an SC model with a small effective range. As it can be seen from the figure, in this case, the short-range SC model does not reproduce the experimental results and seems to be in good agreement with the HD model. In Ref.\cite{Bombin2019}, it was already shown that the HD model is in agreement with the T-matrix results of Ref.~\cite{Schmidt2012}, assuming a short-range model, up to values of the gas parameter $k_\mathrm{F}a_s\sim1$.

\begin{center}
\begin{figure}[tb]
\includegraphics[width=1.02\linewidth]{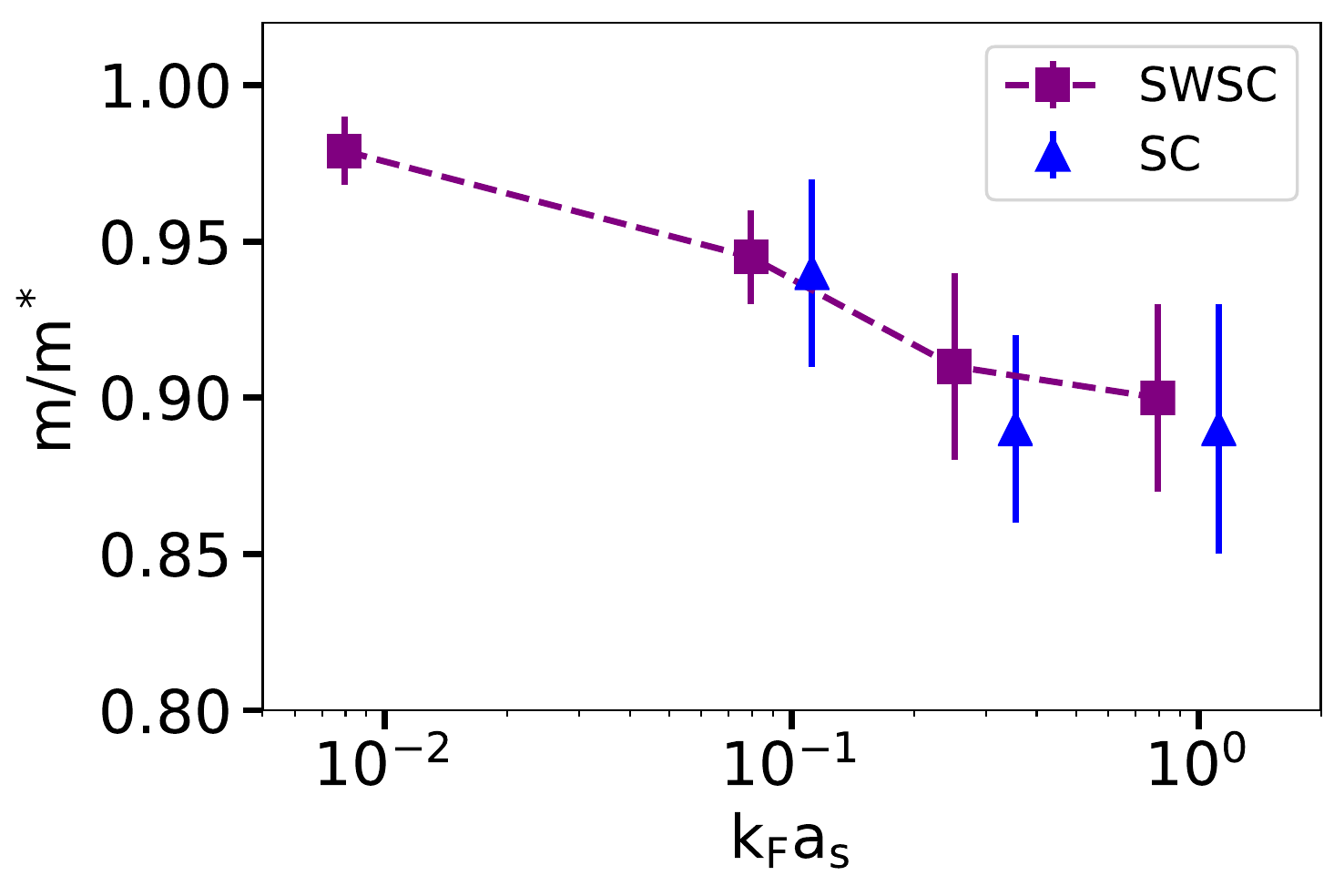}
\caption{Inverse effective mass of the polaron as a function of the the Fermi momentum $k_\mathrm{F}a_s$.
Squares and triangles are results for the SWSC and SC potentials, respectively. The dashed line is a guide to the eye.}
\label{fig:eff_mass}
\end{figure}
\end{center}

In order to provide a further insight into the description of the Fermi polaron as a quasi-particle, we evaluate its effective mass. This mass corresponds to the one of a quasi-particle, formed by the impurity ``dressed'' by the medium, which propagates freely. In the DMC algorithm, the effective mass can be obtained by evaluating the long imaginary-time asymptotic behavior of the diffusion coefficient of the impurity throughout the bath~\cite{Boronat1999,Ardila2015}. 
\begin{equation}
 {m \over m^*} = \lim_{\tau\to\infty} {1\over 4\tau}{D_s^{\downarrow}(\tau) \over D_0} \ ,
\end{equation}
with $D_0 = {\hbar^2 \over 2m}$ corresponding to the free-particle diffusion constant and 
$D_s^{\downarrow}(\tau) = \langle
(\mathbf{r}_\downarrow(\tau)- \mathbf{r}_\downarrow(0))^2\rangle$ the mean squared imaginary-time displacement of the impurity in the medium. The effective mass of the polaron can be experimentally accessed through its low-momenta excitation spectrum,
\begin{equation}
\varepsilon_p(\textbf{k})= \varepsilon_p(\textbf{k}=0) + \frac{\hbar^2}{2m^*} k^2 + \mathcal{O}(k^4)
\end{equation}
with $\varepsilon_p(\textbf{k})$ being the polaron energy corresponding to a state with momentum $\textbf{k}$ and $\varepsilon_p(\textbf{k}=0)$ the ground state polaron energy, as defined in Eq.~\eqref{eq:ener_pol}. Unfortunately, we are not aware of any measurement of $m^*$ in this range of densities to compare with.

We present in Fig. \ref{fig:eff_mass} our DMC results for the effective mass of the polaron. Similarly to previous reported quantities, we show results for the SWSC and the SC models. As can be seen from the figure, both models agree within the statistical error, coming from the MC sampling, for $ k_\mathrm{F}a_s\le 1 $. As expected, the effective mass increases as $k_\mathrm{F}a_s$ approaches to one and the contribution of polaron-medium correlations are enhanced.
However, the observed increase of the effective mass with $k_\mathrm{F}a_s$ is less pronounced that the one predicted by T-matrix theory \cite{Schmidt2012}.

In conclusion, inspired by a recent experiment with $^{173}$Yb \cite{Darkwah2019}, we have addressed the two-dimensional repulsive Fermi polaron problem by means of the DMC technique. The experimental results of Ref.~\cite{Darkwah2019} are outside of the universal regime in terms of the gas parameter~\cite{Bombin2019}. With the aim of reproducing them we include, through the two-body potential, information of the effective range of the impurity-bath interaction. Our results for the polaron energy and the quasi-particle residue show agreement between two different model potentials with the same scattering length and effective range, and are in good agreement with the experimental ones. This hints to the existence of a universal regime in terms of two parameters: the Fermi momentum and the effective range. This assertion seems to be confirmed when the effective mass of the polaron is evaluated. 

Therefore, we have shown that the employment of \textit{ab initio} quantum Monte Carlo techniques can offer insight into the polaron problem once the universality limit is surpassed. 
 Similar results have been shown recently in the formation of droplets in Bose-Bose mixtures~\cite{Cikojevic2019} and in the description of dipolar droplets of Dysprosium atoms~\cite{Bottcher2019}. Besides static properties discussed in this paper, a proper inclusion of the $s$-wave effective range into an underlying density functional might also affect the dynamics \cite{Mukherjee2020induced, Tajima2019collisional,cikojevi2020qmcbased}. We strongly believe that the observation of many-body effects, going beyond the simple mean-field approach,  will stimulate further theoretical and experimental work.

\begin{acknowledgements}
This work has been partly supported by the MINECO (Spain) Grant No. 
FIS2017-84114-C2-1-P and MICINN (Spain) project No. PID2019-107396GB-I00/AEI /10.13039/501100011033. We acknowledge financial support from Secretaría d'Universitats i Recerca del Departament d'Empresa i Coneixement de la Generalitat de Catalunya, co-funded by the European Union Regional Development Fund within the ERDF Operational Program of Catalunya (project QuantumCat, ref. 001-P-001644). 
V. C, acknowledges financial support from the Project HPC-
EUROPA3 (INFRAIA-2016-1-730897), with the support
of the EC Research Innovation Action under the H2020
Programme. J. S-B. acknowledges the FPU fellowship with reference FPU15/01805 from MCIU (Spain).
\end{acknowledgements}

\bibliography{polaron_FR}

\end{document}